\documentclass[amsmath,amssymb,aps,prd,nofootinbib,superscriptaddress,twocolumn]{revtex4-2}
\usepackage[utf8]{inputenc}
\usepackage{mathrsfs}
\usepackage{bm}
\usepackage{url}
\usepackage[normalem]{ulem}
\usepackage{mathtools}
\usepackage{array}
\newcolumntype{P}[1]{>{\centering\arraybackslash}p{#1}}
\newcolumntype{M}[1]{>{\centering\arraybackslash}m{#1}}
\usepackage[caption=false]{subfig}
\usepackage{dcolumn}
\usepackage{graphicx, epsfig}
\usepackage[dvipsnames]{xcolor}
\usepackage{mathrsfs}
\usepackage{bm}
\usepackage{yhmath}
\usepackage[caption=false]{subfig}
\usepackage[normalem]{ulem}
\usepackage{mathtools}
\usepackage{bigints}
\usepackage{float}
\usepackage[colorlinks = true, linkcolor = purple, urlcolor  = blue, citecolor = blue, anchorcolor = blue]{hyperref}
\usepackage{float}
\usepackage{multirow}

\usepackage{tikz,xcolor,hyperref}
\definecolor{darkgreen}{rgb}{0.0, 0.2, 0.13}
\definecolor{bostonuniversityred}{rgb}{0.8, 0.0, 0.0}
\definecolor{lime}{HTML}{A6CE39}
\DeclareRobustCommand{\orcidicon}{
	\begin{tikzpicture}
	\draw[lime, fill=lime] (0,0) 
	circle [radius=0.16] 
	node[white] {{\fontfamily{qag}\selectfont \tiny ID}};
	\draw[white, fill=white] (-0.0625,0.095) 
	circle [radius=0.007];
	\end{tikzpicture}
	\hspace{-2mm}
}
\foreach \x in {A, ..., Z}{\expandafter\xdef\csname orcid\x\endcsname{\noexpand\href{https://orcid.org/\csname orcidauthor\x\endcsname}
			{\noexpand\orcidicon}}
}
\newcommand{\be}{\begin{equation}}
\newcommand{\ee}{\end{equation}}
\newcommand{\ba}{\begin{eqnarray}}
\newcommand{\ea}{\end{eqnarray}}

\newcommand{\nn}{\nonumber}

\newcommand{\half}{\frac{1}{2}}
\def\lp4{$\lambda \phi^4$}
\newcommand{\vv}{{\scriptstyle \mathcal{V}}}

\begin{document}

\title{Cosmological scaling of precursor domain walls}
%
\author{Mainak Mukhopadhyay\hspace{-1mm}\orcidA{}} \email[Corresponding author: ]{mkm7190@psu.edu}
\affiliation{Department of Physics; Department of Astronomy \& Astrophysics; Center for Multimessenger Astrophysics, Institute for Gravitation and the Cosmos, The Pennsylvania State University, University Park, PA 16802, USA
}
\author{Oriol Pujolas\hspace{-1mm}\orcidB{}} \email{pujolas@ifae.es}
\affiliation{Institut de Física d’Altes Energies (IFAE), The Barcelona Institute of Science and Technology (BIST), Campus UAB, 08193 Bellaterra, Barcelona, Spain}

\author{George Zahariade\hspace{-1mm}\orcidC{}} \email{george.zahariade@uj.edu.pl}
\affiliation{Instytut Fizyki Teoretycznej, Uniwersytet Jagiello\'nski, Lojasiewicza 11, 30-348 Krak\'ow, Poland}

\date{\today}

\begin{abstract}
Domain wall (DW) networks have a large impact on cosmology and present interesting dynamics that can be controlled by various scaling regimes. In the first stage after spontaneous breaking of the discrete symmetry, the network is seeded with `DW precursors', the zeros of a tachyonic field. At sufficiently weak coupling, this stage can be quite long. The network is then driven to a non-relativistic scaling regime: in flat spacetime the correlation length grows like $L\sim t^{\kappa}$ with $\kappa=1/2$. 
We focus on the precursor regime in cosmology, assuming a power-law scale factor $a\propto t^\alpha$. We obtain the scaling exponent as a function of the external parameter, $\kappa(\alpha)$, by explicit computation in $1+1$ and $2+1$ dimensions, and
find a smooth transition from nonrelativistic scaling with  $\kappa\simeq 1/2$  for $\alpha\lesssim1/2$ to  DW gas regime $\kappa \simeq \alpha$ for $\alpha\gtrsim1/2$, confirming previous arguments. 
The precise form of the transition $\kappa(\alpha)$ is surprisingly independent of dimension, suggesting that similar results should also be valid in $3+1$ dimensions.
\end{abstract}

\maketitle

\section{Introduction}
\label{sec:intro}
Recent results from pulsar timing arrays \cite{NANOGrav:2023gor,EPTA:2023fyk,EPTA:2023sfo,EPTA:2023xxk,Reardon:2023gzh,Zic:2023gta,Xu:2023wog} reported strong evidence for the existence of a nanohertz (nHz) stochastic gravitational wave (GW) background. Although the most widely accepted interpretation of this evidence is a GW background generated by supermassive black hole binary systems, other possibilities and sources are reasonably plausible, including primordial GW backgrounds originating from beyond standard model (BSM) physics. In particular, a GW background from cosmic domain walls (DWs) could be an alternative interpretation of the signal  \cite{Bian:2020urb,Sakharov:2021dim,Bian:2022tju,Ferreira:2022zzo,NANOGrav:2023hvm,Blasi:2023sej,Ellis:2023oxs,Ferreira:2024eru}.

DWs are codimension one topological defects that are generically formed as a result of the spontaneous breaking of a discrete symmetry, see \cite{Kibble:1976sj,Vilenkin:2000jqa,Vachaspati:2006zz,Saikawa:2017hiv} for reviews. They are stable solutions of classical field theories with a discrete vacuum manifold, that smoothly interpolate between spatial regions with different vacua. Various BSM scenarios, including QCD axion models \cite{Sikivie:1982qv,Vilenkin:1982ks}, lead to  the formation of such defects in the early universe. The energy density stored in such objects decays more slowly than matter leading to a so-called \textit{DW problem} \cite{Zeldovich:1974uw,Sikivie:1982qv,Vilenkin:2000jqa}. There are natural mechanisms to avoid DW domination so the problem in reality is a virtue: the tendency of DWs to dominate the energy density implies that models with DWs leave a large impact on cosmology. Phenomenologically, this can be in the form of primordial black holes, GWs, and/or other relics.

In order to confront the DW models with observations it is important to know in detail their dynamics, abundance etc. As is well known, the evolution of a DW network is dominated by a self-similar `scaling' attractor, characterized by having about one Hubble-sized DW per Hubble volume at all times. Most of the previous work on DW evolution result from numerical simulations in classical field theory simulations \cite{Press:1989yh,Larsson:1996sp,Garagounis:2002kt,Oliveira:2004he,Hiramatsu:2010yz,Leite:2011sc,Kawasaki:2011vv,Hiramatsu:2013qaa,Correia:2014kqa,Correia:2018tty,ZambujalFerreira:2021cte,Kitajima:2023cek,Chang:2023rll,Ferreira:2024eru}. 
In the analytical front, results derived using the thin wall limit (combined with a mean field approximation) \cite{Hindmarsh:1996xv,Hindmarsh:2002bq} are able to reproduce some scaling properties of DW networks in cosmology. 

Further analytic progress was made in~\cite{Mukhopadhyay:2020gwc,Mukhopadhyay:2020xmy} and \cite{Pujolas:2022qvs} by addressing the problem using quantum field theory methods. The formation of a DW network upon $\mathbb{Z}_2$ symmetry breaking transition was analyzed exploiting the \textit{DW precursor} limit, where the symmetry is broken but the field hasn't stabilized yet. This already leads to well defined dynamical DW objects (the seeds of the actual DWs formed when the field is nonliear) that arise in the linear theory and so their behaviour can be computed.

Interestingly, a network of precursor DWs formed after the $\mathbb{Z}_2$ quench reaches a scaling regime
characterized by a correlation length that grows in time like $L\sim t^{1/2}$ \cite{Mukhopadhyay:2020gwc,Mukhopadhyay:2020xmy}, see also \cite{Pujolas:2022qvs}. 
These works assumed the DW precursor limit and a Minkowski background. In this paper we aim to relax the latter assumption and extend the previous analysis to an expanding spacetime more directly relevant to cosmology. 

In Ref.~\cite{Pujolas:2022qvs}, a simple argument was given  on how the $t^{1/2}$ scaling of precursor DW networks extends to cosmology. The origin of the  $t^{1/2}$ scaling resides in the fact that precursor DWs are non-relativistic at late times. This is expected to hold in a slowly expanding background, and the scaling should continue to be $t^{1/2}$ in terms of cosmic time. However, this cannot remain true in cosmologies with a sufficiently fast expansion rate. For power law cosmologies, $a(t)\sim t^\alpha$, with $\alpha \gtrsim 1/2$ one expects a change of behaviour to a DW gas regime where $L(t)\propto a(t)$. From this argument one cannot tell whether this transition is sharp, occurring at the precise value $\alpha=1/2$, or rather smooth, and to what extent. Also, $\alpha=1/2$ should mark the onset transition independently of $d$, the number of space dimensions, although the details of the transition might depend on $d$.
The purpose of this work is to exhibit this transition in detail by direct calculation.

For simplicity we first work in $1+1$ dimensions, and thus limit ourselves to studying the evolution of the number density of kinks (and antikinks) following a $\mathbb{Z}_2$-symmetry breaking phase transition in an expanding background. We assume a spatially flat Friedmann-Lema\^itre-Robertson-Walker (FLRW) background metric with a scale factor that grows as a power law with respect to cosmological time. In particular, we aim to understand how the growth of the scale factor affects the late time scaling of the kink number density. We then extend the results to domain walls in $2+1$ dimensions.

The paper is organized as follows. We begin by reviewing the formalism and setting up the main equations for the $1+1$ dimensional case in Sec.~\ref{sec:kinks} (with more details included in the appendices). The extension to the $2+1$ dimensional case is briefly outlined in   Sec.~\ref{sec:dws}. Our main results are showcased in Sec.~\ref{sec:res}. We discuss the effective dynamics of precursor DWs in the velocity-one-scale (VOS) model in Sec.~\ref{sec:VOS}. We conclude with a brief discussion in Sec.~\ref{sec:disc}.
\section{Kinks in an expanding background}
\label{sec:kinks}
In this section we extend the formalism of Refs.~\cite{Mukhopadhyay:2020gwc,Mukhopadhyay:2020xmy,Pujolas:2022qvs} to a $1+1$ dimensional expanding spacetime. As usual, dotted quantities denote time derivatives while primed quantities denote spatial derivatives. We work in units where $\hbar = c = 1$.

The aim is to investigate the dynamics of the average number density of kinks formed as a scalar field $\phi(t,x)$ undergoes a quantum phase transition\footnote{We consider a quantum phase transition implying a zero temperature vacuum state for the field. However, we expect that our results on the late time scaling would also apply to a thermal transition~\cite{CALZETTA198932,Boyanovsky:1993fy,Ibaceta:1998yy,Boyanovsky:1999wd}. This is because the precursor DWs are over-damped and hence their late time behaviour follows non-relativistic scaling \cite{Pujolas:2022qvs} (refer to Sec.~\ref{sec:VOS} for more details).
}. The background  metric is taken to be
$ds^2=dt^2-a(t)^2 dx^2$, with scale factor
\be
\label{eq:scfac}
a(t) \propto t^\alpha\,,
\ee
where $0\leq\alpha<1$ to focus on the case of decelerating expansion. The action for the scalar field is thus given by
\be
\label{eq:action}
S = \int dt\, dx\,a(t)\left[\frac{1}{2}\dot{\phi}^2-\frac{1}{2}a(t)^{-2}\phi^{\prime 2}-\frac{1}{2}m^2(t)\phi^2+\cdots\right],
\ee
where 
\be
\label{eq:m2}
m^2 (t) = -m^2 {\rm tanh} \bigg( \frac{t - t_{\rm pt}}{\tau} \bigg),
\ee
and the dots denote stabilizing terms for the potential e.g. $\lambda \phi^4$. In this work (as in~\cite{Mukhopadhyay:2020gwc,Mukhopadhyay:2020xmy,Pujolas:2022qvs}) we neglect such terms by focusing on the so-called {\it spinodal instability} phase which can last for arbitrary long time if the self-coupling $\lambda$ is weak enough. The phase transition is triggered at $t=t_{\rm pt}>0$ when the mass squared becomes negative and the potential turns tachyonic. The {\it quench time scale} $\tau<t_{\rm pt}$ parametrizes the rapidity of the phase transition. Notice that we expect kinks (and antikinks) to form as a consequence of this phase transition, but since we neglect self-coupling terms we can only strictly talk about kink (and antikink) precursors. Therefore, and in order to not overburden the reader, we will assume the word precursor to be implied even when not explicitly stated.

Let us now summarize the task at hand. We want to determine the average number density of kinks formed during the phase transition as a function of time, $n_K(t)$. This roughly boils down to counting \emph{zeros} of the scalar field $\phi$. Since we are interested in finding the quantum dynamics of $n_K(t)$ we will need to compute the quantum average of a suitably constructed {\it number density of zeros} operator in a state $|\Psi\rangle$ of the quantized field $\phi$. We will work in the Schr\"odinger picture and choose the state $|\Psi(t)\rangle$ to coincide with the zeroth order adiabatic vacuum state of the quadratic theory at some time $t_0$ verifying $0<t_0< t_{\rm pt}$. This will depend only on the values $m^2(t_0)$ and $a_0\equiv a(t_0)$. If $t_{\rm pt}-t_0\gg\tau$, our physical results will not depend on the choice of initial time since $m^2(t_0)\approx m^2$ does not depend on $t_0$ while $a_0$ only serves to rescale $L$ and thus has no impact in the infinite volume limit.

By discretizing the field theory on a periodic lattice of size $L$ with $N$ evenly spaced points (such that the lattice spacing is given by $\ell \equiv L/N$) and deploying the methods of Refs.~\cite{Mukhopadhyay:2020gwc,Mukhopadhyay:2020xmy,Pujolas:2022qvs} (see Appendix~\ref{appsec:derivation}) we find the probability density functional for a given discretized field configuration $\bm{\phi}\equiv (\phi_1, \phi_2, \dots, \phi_N)^{T}$ at time $t$ to be\footnote{Here we use the notation $\phi_i \equiv \phi(t, i \ell)$ for the discretized field values, and we denote the transpose using $^T$.}
\be
\label{eq:probden}
\mathcal{P} (t;\bm \phi) = \frac{1}{\sqrt{{\rm det} \big(2 \pi \bm K(t)\big)}} e^{-\bm \phi^T \bm K(t)^{-1} \bm \phi/2},
\ee
where the covariance matrix $\bm K(t)$ is given by 
\be
\label{eq:green}
[\bm K(t)]_{jk} = \frac{1}{N} \sum_{n = 0}^{N-1} |c_n(t)|^2 {\rm cos} \big( 2 \pi n (j-k)/N \big),
\ee
and the mode functions $c_n(t)$ (for $0\leq n<N$) obey 
\be
\label{eq:cn}
\ddot{c}_n + H(t) \dot{c}_n + \left[ \frac{4}{(a(t)\ell)^2} {\rm sin}^2 \left( \frac{\pi n}{N} \right) + m^2(t) \right] c_n = 0,
\ee
$H\equiv \dot{a}/a$ being the Hubble rate. Our particular choice of quantum state imposes the specific initial conditions
\ba
c_n (t_0) &=& 
\frac{1}{\sqrt{2a_{0}\ell}} 
\left [  \frac{4}{(a_0\ell)^2}\sin^2 \left ( \frac{\pi n}{N}\right ) + m^2(t_0) \right ]^{-1/4},
\label{eq:cn0}\\
{\dot c}_n (t_{0}) &=& \frac{i}{\sqrt{2a_{0}\ell}} 
\left [ \frac{4}{(a_0\ell)^2}\sin^2 \left (\frac{\pi n}{N} \right ) + m^2(t_{0}) \right ]^{1/4}.
\label{eq:cndot0}
\ea
Since we effectively work in a regime where the Gaussian approximation holds, the mode functions are seen to contain all the information about the quantum dynamics of the system.

The quantum operator used to estimate  the average kink number density is the direct generalization of the flat spacetime one described at length in~\cite{Mukhopadhyay:2020gwc,Mukhopadhyay:2020xmy,Pujolas:2022qvs}. Essentially, the idea is to look for kinks and antikinks among the zeros of the field configuration. Of course quantum fluctuations introduce spurious zeros that need to be discarded and we will do so through a prescription which will be briefly described later on. We define a {\it number density of zeros} operator by
\be
n_{\rm Z}\left(\hat{\bm\phi}\right)\equiv\frac{1}{a(t) L}\sum_{i=1}^N\frac{1}{4}\left[{\rm sgn}\left(\hat{\phi}_i \right)-{\rm sgn}\left(\hat{\phi}_{i+1} \right)\right]^2,
\ee
where the symbol ${\rm sgn}$ denotes the {\it signum} function. This counts the number density of sign changes of the $\phi$ field profile within the (comoving) size $a(t)L$ of the lattice, to a level of accuracy determined by its coarseness $\ell$.

Using the translational invariance of the problem, the vacuum expectation value of this operator in the state $\left|\Psi\right\rangle$ can readily be computed (see~\cite{Mukhopadhyay:2020gwc,Mukhopadhyay:2020xmy,Pujolas:2022qvs} for details). It reads
\ba
\label{finalxnz}
\left\langle n_{\rm Z}\left(\hat{\bm\phi}\right)\right\rangle &\equiv& \int d\phi_1\cdots d\phi_n\, n_{\rm Z}\left(\hat{\bm\phi}\right)\mathcal{P} (t;\bm \phi) \nn\\
&=&\frac{N}{\pi a(t) L}\cos^{-1}\left(\frac{\beta}{\alpha}\right),
\ea
where $\alpha$ and $\beta$ are given by
\ba
\alpha & = & \frac{1}{N} \sum_{n=1}^N |c_n|^2, \label{eq:alpha} \\
\beta & = & \frac{1}{N} \sum_{n=1}^N |c_n|^2 \cos (2\pi n/N).\label{eq:beta}
\ea

As we have already mentioned, we expect this quantity to not be an accurate estimator of the number density of kinks and antikinks (even for very fine lattices) because of the random fluctuations of the field caused by oscillating modes. This can remediated by restricting the sums in Eqs.~\eqref{eq:alpha} and \eqref{eq:beta} to only those modes $c_n(t)$ that are \emph{non-oscillating} or in other words \emph{unstable}. Such modes verify
\be
\label{eq:omega2n}
\omega_n^2(t)\equiv\frac{4}{(a(t)\ell)^2}\sin^2 \left (\frac{\pi n}{N} \right ) + m^2(t) \leq 0.
\ee
Before the phase transition, that is, for $t<t_{\rm pt}$, all the modes are stable and there are no kinks or antikinks. For $t>t_{\rm pt}$ the time-dependent critical value of $n$, $n_c(t)$, separating the stable modes from the unstable ones is given by
\be
\label{eq:ncrit}
n_c(t)\equiv \left\lfloor\frac{N}{\pi}\sin^{-1}\left(\frac{a(t) \ell \sqrt{|m^2(t)|}}{2}\right)\right\rfloor,
\ee
where $\lfloor \cdot\rfloor$ denotes the integer part function.
Thus we find the average number density of kinks $n_K$ to be
\be
\label{eq:nk}
n_K = \frac{N}{\pi a(t) L} {\rm cos}^{-1} \left( \frac{\bar{\beta}}{\bar{\alpha}} \right),
\ee
where
\ba
\bar{\alpha} &\equiv& \frac{1}{N} \sum_{\omega_n^2(t)\leq 0} |c_n|^2 = \frac{1}{N} \left[ |c_0|^2 + 2 \sum_{n=1}^{n_c(t)} |c_n|^2 \right],
\label{baralpha} \\
\bar{\beta} &\equiv& \frac{1}{N} \sum_{\omega_n^2(t)\leq 0} |c_n|^2 \cos (2\pi n/N) \nn\\
&=& \frac{1}{N} \left[ |c_0|^2+ 2\sum_{n=1}^{n_c(t)} |c_n|^2 \cos (2\pi n/N) \right].
\label{barbeta}
\ea
Notice that the number of unstable modes grows with time (after the phase transition). This is due to the physical mode wavelengths getting stretched by the (cosmological) expansion of the lattice or in other words to UV modes becoming IR. In fact, for fixed lattice spacing $\ell$, after a time of order $t/t_0\sim (a_0\ell)^{-1/\alpha}m^{-1/\alpha}$ all modes are unstable and the estimate of $n_K(t)$ ceases to be accurate: the longer we want to track its dynamics, the more modes we will need to take into consideration (and the finer our lattice needs to be). 

In general, Eq.~\eqref{eq:cn} will need to be solved numerically allowing us to compute $\bar{\alpha}$, $\bar{\beta}$, and finally $n_K$ via Eq.~\eqref{eq:nk}. Our results will be showcased in Sec.~\ref{sec:res} and we leave the more precise algorithmic details to Appendix~\ref{appsubsec:numerics_equations}.

Before ending this section, let us quickly mention that although it is not possible to estimate $n_K$ analytically in general, or for arbitrary rates of expansion $\alpha$ and phase transition time scales $\tau$, the case of the sudden phase transition $\tau=0$ constitutes a notable exception. In this scenario $m^2(t) = -m^2 \Theta(t-t_{\rm pt})$, where $\Theta$ denotes the Heaviside step function, and one can choose $t_{0} = t_{\rm pt}^-$. Equation~\eqref{eq:cn} then reduces to
\be
\label{eq:inst_cn}
\ddot{c}_n + \frac{\alpha}{t} \dot{c}_n + \left[ \frac{4}{(a_0 \ell)^2 (t/t_{\rm pt})^{2 \alpha}} \sin^2 \left( \frac{\pi n}{N} \right) - m^2 \right] c_n = 0,
\ee
with initial conditions,
\ba
\label{eq:cn_init}
c_n (t_{\rm pt}) &=& 
\frac{1}{\sqrt{2a_0 \ell }} 
\left [ \frac{4}{(a_0\ell)^2}\sin^2 \left ( \frac{\pi n}{N}\right ) + m^2 \right ]^{-1/4},
\label{ckt0sudden}\\
{\dot c}_n (t_{\rm pt}) &=& \frac{i}{\sqrt{2a_0 \ell}}  
\left [ \frac{4}{(a_0\ell)^2}\sin^2 \left (\frac{\pi n}{N} \right ) + m^2 \right ]^{1/4}.
\label{dotckt0sudden}
\ea
It is still non-trivial to find a closed form expression for $c_n(t)$ for general $\alpha$. However, the particular cases of $\alpha = 0$ and $1$ can be solved analytically in the continuum ($\ell\rightarrow 0$) and infinite volume ($L\rightarrow\infty$) limit\footnote{In the $\alpha=1/2$ case, Eq.~\eqref{eq:inst_cn} takes the form of a hypergeometric differential equation which can be solved analytically as well but it is no easy task to use the resulting closed-form expressions for a useful estimate of $n_K(t)$.}, thus giving access to closed form expressions for $n_K$ and providing in principle a useful benchmark against which to gauge the accuracy of our numerics. The case of Minkowski spacetime, $\alpha = 0$, has already been studied in detail in Refs.~\cite{Mukhopadhyay:2020gwc,Mukhopadhyay:2020xmy,Pujolas:2022qvs} and it was found that $n_K(t) \sim t^{-1/2}$ for late times. In the case of linear expansion, $\alpha = 1$, Eq.~\eqref{eq:inst_cn} takes the form of a Bessel differential equation whose solutions are given in terms of modified Bessel functions $I_n$ and $K_n$,
\ba
c_n(t) &=& \frac{1}{2 \sqrt{2\ell} (k_n^2 + m^2)^{1/4} }\left[ I_{ik_n}(mt) \left( m \left( K_{ik_n - 1}(m)+ \right. \right. \right. \nn\\ 
&& \left. \left. \left. \hspace{-1cm} K_{ik_n+1}(m) \right) + 2i \sqrt{k_n^2 + m^2} K_{ik_n}(m) \right) +2 K_{ik_n} (mt)\right.\nn\\
&&\left. \hspace{-1cm} \left( m I_{ik_n+1}(m)- i \left( \sqrt{k_n^2 + m^2} - k_n\right) I_{ik_n}(m) \right)  \right],
\ea
where $k_n \equiv 2\sin(\pi n/N)/\ell$ and we chose $a_0 = 1$, $t_{\rm pt} = 1$ to simplify the expressions. In the late time limit $I_{z}(mt)\sim e^{mt}/\sqrt{2\pi mt}$ while $K_z(mt)\sim \sqrt{2}\, e^{-mt}/\sqrt{\pi mt}$, for all $z\in\mathbb{C}$ (see e.g. p.203 in~\cite{watson1944treatise}, (10.40.5) in~\cite{822801}), so it is easy to see that the time dependence in $|c_n(t)|^2$ factors out and is equal to $e^{2mt}/t$. Therefore the ratio $\bar{\beta}/\bar{\alpha}$ becomes time independent at large times and $n_K(t)\sim 1/a(t)\sim t^{-1}$. This corresponds to the domain wall gas limit and we will discuss it more in Sec.~\ref{sec:res}.

The previous discussion applied to kinks and antikinks in one spatial dimension. Such objects are expected to behave like point particles and do not have any internal structure, so we now proceed to briefly extend the formalism to similar but more complex objects: domain walls in two (and higher) spatial dimensions.
\section{Domain walls}
\label{sec:dws}
The formalism outlined in Sec.~\ref{sec:kinks} for kinks in $1+1$ dimensions can be trivially extended to study the dynamics of the {\it domain wall length density} in a $2+1$ dimensional theory with a real scalar field $\phi(t,x,y)$ undergoing a $\mathbb{Z}_2$-symmetry breaking phase transition. We begin with the Lagrangian density given by
\ba
\label{highd_lagrangian}
S&=&\int dt\, dx\, dy\,a^2(t)\left[ \half (\partial_t\phi)^2-\half a^{-2}(t)(\partial_x\phi)^2 \right.\nn\\ 
&&\hspace{1cm} \left. -\half a^{-2}(t) (\partial_y\phi)^2 
- \half m^2(t) \phi^2+\cdots\right],
\ea
where the definitions of $m^2(t)$ and $a(t)$ are unchanged, and we discard any non-linear stabilizing terms.
Repeating the steps outlined in the previous section (and carefully dealing with formal complications due to the presence of two spatial indices), it is not hard to convince oneself that the quantum dynamics of the system is fully determined by the mode functions $c_{n,n'}(t)$ (for $0\leq n,n'<N$) verifying
\ba
\label{eq:cn_dw}
\ddot{c}_{n,n'} &+& 2 H(t) \dot{c}_{n,n'} + \left[ \frac{4}{(a(t)\ell)^2} \left\{ {\rm sin}^2 \left( \frac{\pi n}{N} \right) \right. \right. \nonumber\\
&&+ \left. \left. {\rm sin}^2 \left( \frac{\pi n'}{N} \right)  \right\} + m^2(t) \right] c_{n,n'} = 0,
\ea
with initial conditions
\ba
\label{eq:cndw_init}
c_{n,n'}(t_0) &=& 
\frac{1}{a_{0} \ell \sqrt{2}} 
\left[  \frac{4}{(a_0\ell)^2} \left\{ \sin^2 \left ( \frac{\pi n}{N}\right ) \right. \right. \nonumber \\
&&+ \left. \left. \sin^2 \left ( \frac{\pi n'}{N}\right ) \right\} + m^2(t_0) \right ]^{-1/4},
\label{ckt0_dw}\\
{\dot c}_{n,n'} (t_{0}) &=& \frac{i}{a_{0}\ell \sqrt{2}} 
\left [ \frac{4}{(a_0\ell)^2}  \left\{ \sin^2 \left (\frac{\pi n}{N} \right ) \right. \right. \nonumber \\
&&+ \left. \left. \sin^2 \left (\frac{\pi n'}{N} \right ) \right\} + m^2(t_{0}) \right ]^{1/4}.
\label{dotckt0_dw}
\ea

As mentioned above the relevant quantity that needs to be computed here is the domain wall length density (in three spatial dimensions this corresponds to the more familiar {\it domain wall area density}). As discussed in detail in Ref.~\cite{Pujolas:2022qvs}, this is obtained from the quantum average of a {\it length density of zeros} operator\footnote{Again we use the notation $\phi_{ij} \equiv \phi(t, i \ell,j \ell)$ for the discretized field values.}
\ba
\mu_Z\left(\hat{\bm{\phi}}\right)&\equiv&\frac{a(t)\ell}{(a(t)L)^2}\sum_{i,j=1}^N\frac{1}{4}\left[{\rm sgn}(\hat{\phi}_{ij})-{\rm sgn}(\hat{\phi}_{i+1,j})\right]^2\nn\\
&&\hspace{-1cm}+\frac{a(t)\ell}{(a(t)L)^2}\sum_{i,j=1}^N\frac{1}{4}\left[{\rm sgn}(\hat{\phi}_{ij})-{\rm sgn}(\hat{\phi}_{i,j+1})\right]^2.
\ea
Notice that, for small enough $\ell$, a domain wall can cross a unit cell in the $x$ or in the $y$ direction, hence the presence of two distinct terms in the above definition. While this is of course an idealization, since domain walls can have arbitrary orientations, we expect the error to be $\mathcal{O}(1)$ on average.

Using the translational and rotational invariance of the problem we find\footnote{Note that in Ref.~\cite{Pujolas:2022qvs} there is a typo in Eq.~(59), where a factor of $2$ appears in the denominator of formula for the average DW area density (instead of the numerator). This however does not alter the main results.}
\be
\label{eq:ndw}
\left\langle\mu_{Z}\left(\hat{\bm{\phi}}\right)\right\rangle= \frac{2N}{\pi a(t) L} {\rm cos}^{-1} \left( \frac{{\beta}}{{\alpha}} \right)
\ee
where
\ba
\alpha(t)&\equiv& \frac{1}{N^2}\sum_{n,n'=0}^{N-1}|c_{n,n'}(t)|^{2},\\
\beta(t)&\equiv& \frac{1}{N^2}\sum_{n,n'=0}^{N-1}|c_{n,n'}(t)|^{2} \cos(2\pi n'/N).
\ea
The factor $2$ appearing in~\eqref{eq:ndw} is a consequence of the higher spatial dimension. 
Once again the actual domain wall length density, $\mu_{DW}$, is obtained from~\eqref{eq:ndw} by requiring the sums in $\alpha$, $\beta$ to run only over those modes that are unstable. Numerical results for the dynamics of the domain wall length density in two spatial dimensions will be presented in the next section.

It is also easy to see how the formalism can be extended to higher spatial dimensions $d$, where the quantum average of the {\it area density of zeros} of the field $\phi$ is obtained by replacing the factor $2$ in~\eqref{eq:ndw} by $d$ and $\alpha$, $\beta$ are defined in an analogous way with the help of the higher dimensional mode functions.

\section{Results}
\label{sec:res}
We are now in a position to discuss the main results of this work. We begin with the one spatial dimension case of Sec.~\ref{sec:kinks}. The dynamics of the kink number density $n_K(t)$ is obtained by determining the mode functions $c_n(t)$ and plugging them in Eq.~\eqref{eq:nk}. More precisely we numerically solve the differential equations~\eqref{eq:cn} with initial conditions given in~\eqref{eq:cn0} and~\eqref{eq:cndot0}\footnote{For numerical reasons we solve a slightly different but equivalent set of equations, see Appendix~\ref{appsubsec:numerics_equations} for details.} using an explicit Crank-Nicholson method. The computations were performed using a \emph{parallelized CPU code}. Working in units where $m=1$ (and unless specified otherwise) we set the time of the phase transition to $t_{\rm pt} = 1$ and start evolving the equations at $t_0 = 0.5$. To have an accurate estimate of the average kink number density up to $\mathcal{O}(10^3)$ times for all values of $\alpha$ we need to choose the parameters $L, N,$ and the time increment of the numerical evolution method, $dt$, carefully. We discuss this in detail in Appendix~\ref{appsubsec:numerics_2}. 

\begin{figure}[h]
\centering
\includegraphics[width=0.48\textwidth]{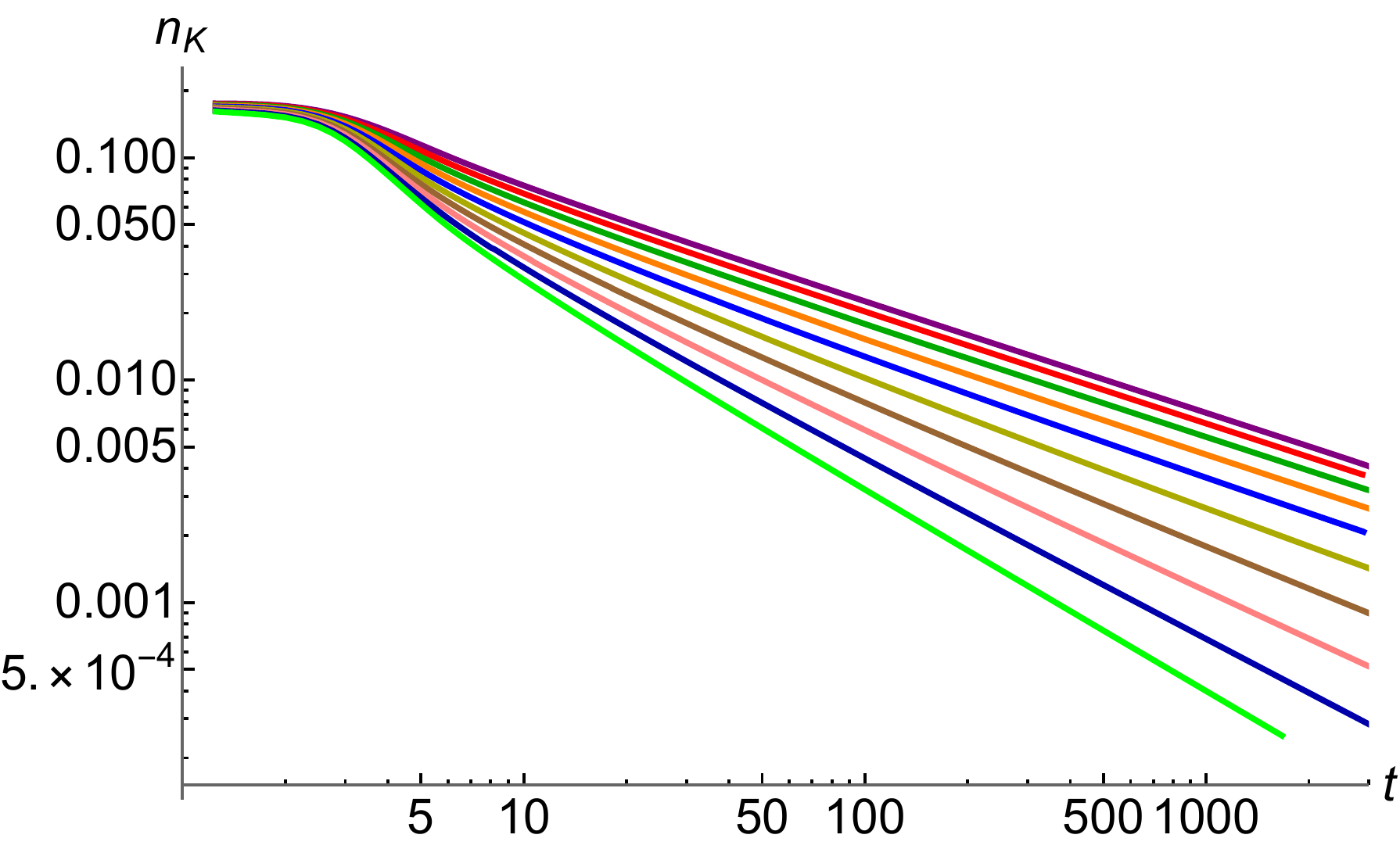}
\caption{\label{fig:numden_pow} Log-log plot of the average number density of kinks $n_K$ as a function of time for different values of the power law index $\alpha$. The power law indices are color coded as follows: $\alpha = 0$ (purple), $\alpha = 0.1$ (red), $\alpha = 0.2$ (darker green), $\alpha = 0.3$ (orange), $\alpha = 0.4$ (blue), $\alpha = 0.5$ (darker yellow), $\alpha = 0.6$ (brown), $\alpha = 0.7$ (pink), $\alpha = 0.8$ (darker blue), and $\alpha = 0.9$ (green). The parameters used are $t_{\rm pt} = 1$, $t_0 = 0.5$, $\tau = 0.01$, and $a_0 = 1$ (in units where $m=1$). For details on the choice of $L$, $N$, and $dt$, see Appendix~\ref{appsubsec:numerics_2}.
}
\end{figure}
In Fig.~\ref{fig:numden_pow}, we show the average number density of kinks $n_K(t)$ as a function of time $t$ for different values of the index $\alpha$ associated with the power law growth of the scale factor~\eqref{eq:scfac}. The phase transition occurs at $t_{\rm pt} = 1$, before which $n_K = 0$.  The quench timescale is taken to be $\tau = 0.01$, which makes the phase transition quasi-instantaneous. Thus immediately after the phase transition kinks and antikinks are quickly produced with random positions and velocities and $n_K$ reaches a maximum value almost instantaneously. Then, as kink/antikink annihilation becomes important, $n_K$ begins to decrease. At late times, similarly to Refs.~\cite{Mukhopadhyay:2020gwc,Mukhopadhyay:2020xmy}, we observe a power law decline with $n_K(t) \propto t^{-\kappa}$. For $\alpha = 0$ (Minkowski spacetime) we reproduce the results from~\cite{Mukhopadhyay:2020gwc,Mukhopadhyay:2020xmy} where we see that at late times $n_K (t) \propto t^{-1/2}$, or $\kappa = 0.5$. As the value of $\alpha$ increases, the slope associated with the late time power law decay of $n_{K}$ also increases. This is due to the fact that the expansion helps dilute the leftover kinks and antikinks. 

\begin{figure}
\centering
\includegraphics[width=0.48\textwidth]{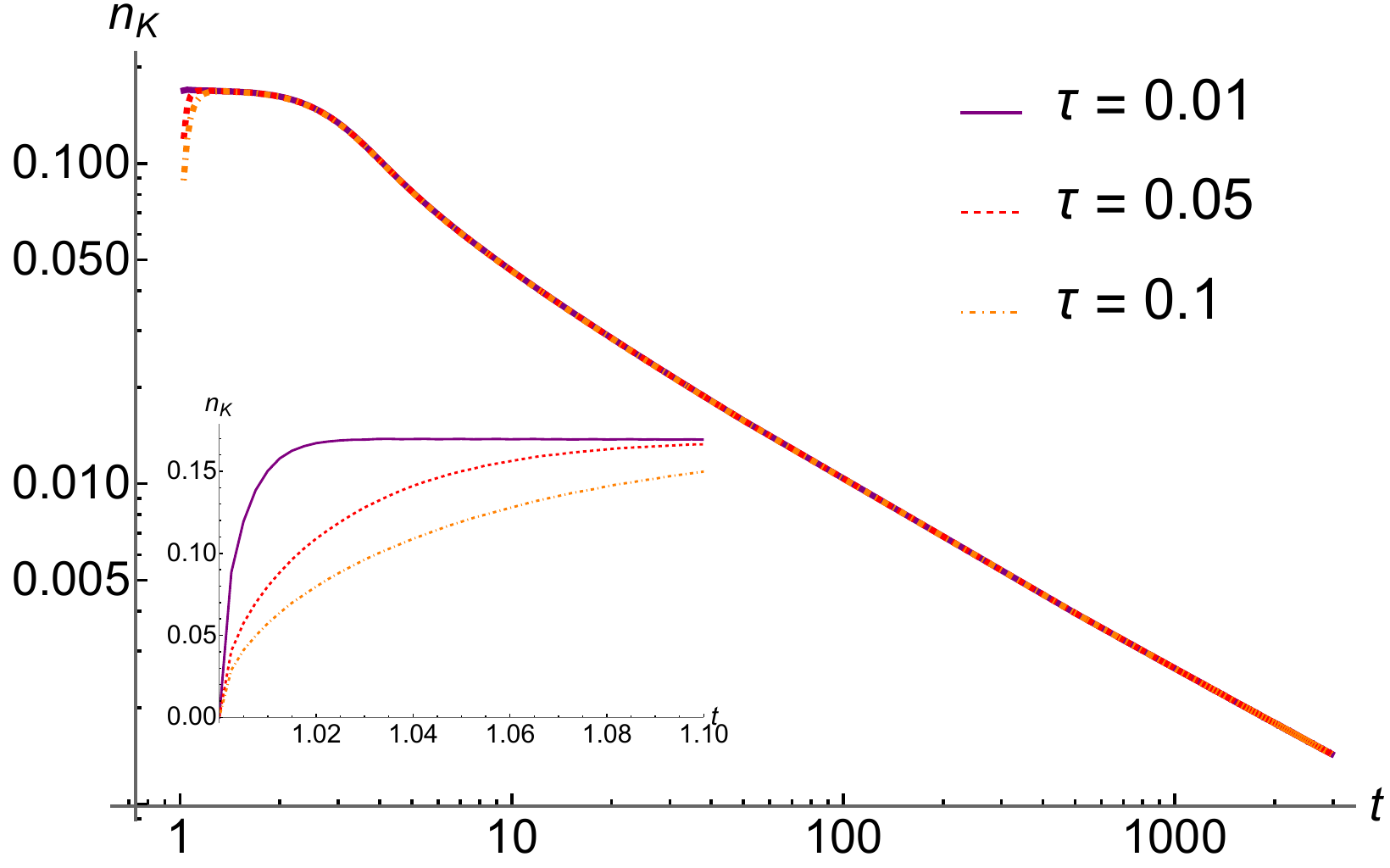}
\caption{\label{fig:numden_tau_full} Log-log plot of the average kink number density $n_K$ as a function of time for different values of the quench timescale $\tau = 0.01$ (solid purple), $0.05$ (dashed red), and $0.1$ (dot-dashed orange). The inset provides a zoomed in view of the plot at early times after the phase transition. The parameters used are $t_{\rm pt} = 1$, $t_0 = 0.5$, $\alpha = 0.5$, $L = 32,\ N = 12800,\ dt=0.25$, $a_0 = 1$ (in units where $m=1$). The inset is made using a finer temporal step, $dt = 0.0025$ (see Appendix~\ref{appsubsec:numerics_2}).
}
\end{figure}
In Fig.~\ref{fig:numden_tau_full} we show $n_K$ as a function of time for different values of the quench timescale $\tau$ ($0.01$, $0.05$ and $0.1$) and $\alpha =0.5$. We can see that at late times the curves for $n_K(t)$ overlap, which proves the universality of the power law decay (or its insensitivity to the details of the phase transition). However the growth of $n_K$ immediately after the phase transition is different for the three cases as can be seen from the inset in Fig.~\ref{fig:numden_tau_full}. We note that as $\tau$ increases the growth rate of $n_K$ decreases. (In fact the maximum value of $n_K$ also decreases but this is not clearly visible on the plot.) This implies that the quicker the phase transition (shorter quench timescales) the more efficient the kink production rate is.
\begin{figure}
\centering
\includegraphics[width=0.49\textwidth]{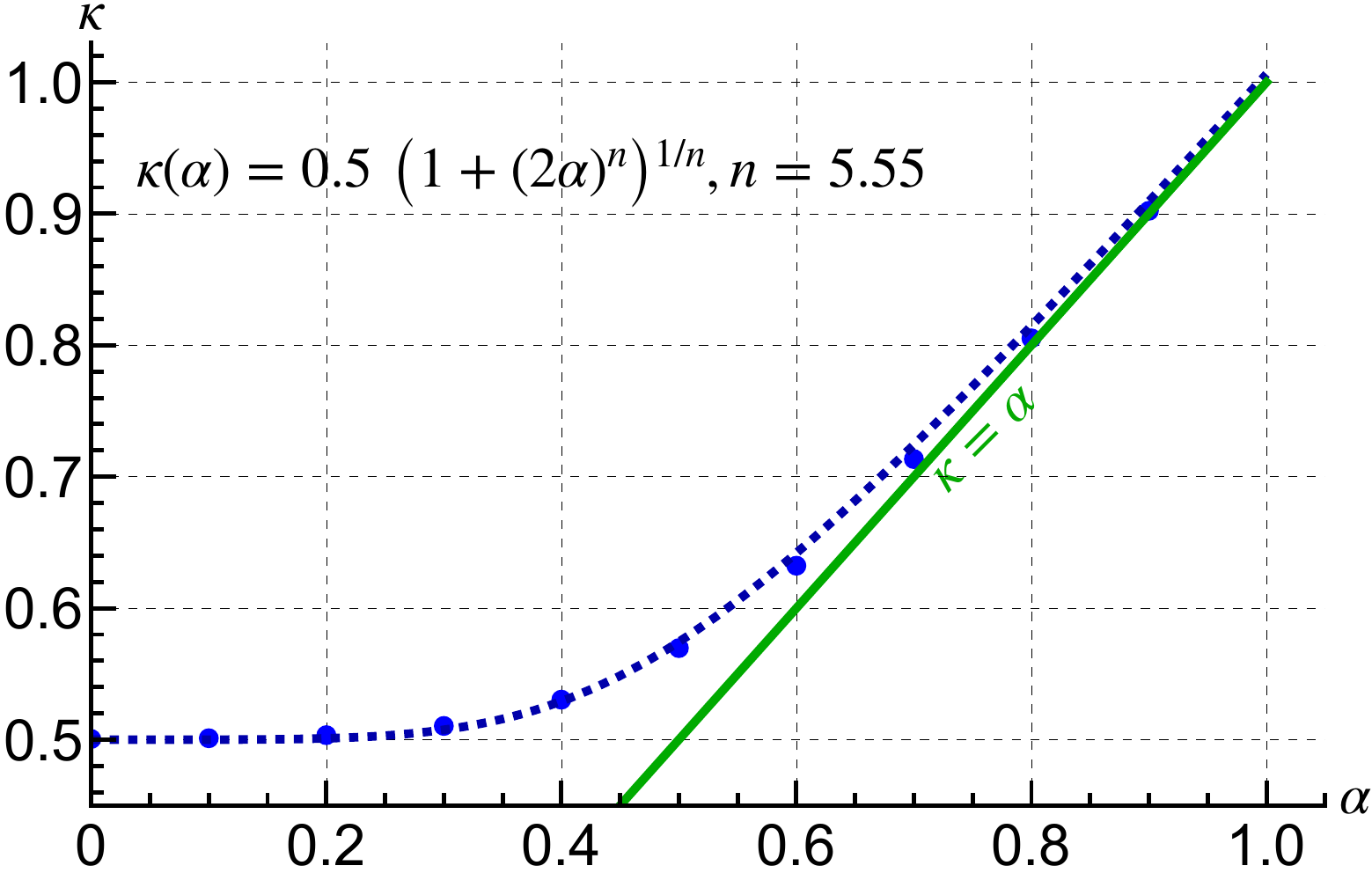}
\caption{\label{fig:lambda_alpha} The late time power law index $n_K(t)$, $\kappa$, given by $n_K(t)\sim t^{-\kappa}$ as a function of the power law index, $\alpha$,  associated with the growth of the scale factor (see Eq.~\eqref{eq:scfac}). The dashed dark blue line shows the broken linear function fit to the data points. For $\alpha \lesssim 1$ the dashed line has a slope of $\sim 1$, that is, $\kappa \sim \alpha$. The solid green line shows $\kappa = \alpha$. For details on the choice of $L$, $N$, and $dt$, see Appendix~\ref{appsubsec:numerics_2}.
}
\end{figure}

To make the relationship between the expansion rate and the power law followed by $n_K$ at late times more explicit, we also plot the power law index $\kappa$ versus $\alpha$ in Fig.~\ref{fig:lambda_alpha}. As $\alpha$ varies from $0$ to $1$, $\kappa$ smoothly interpolates between $0.5$ and a linear function with slope $1$. This agrees with the analytic estimate of the previous section. For $\alpha \lesssim 1$, $\kappa \sim \alpha$ which means that the average number of kinks and antikinks per comoving volume remains approximately constant and they are unable to mutually annihilate because of the Hubble flow: they are frozen and behave like a dilute non-interacting gas. Any further dilution is entirely due to the increase of the physical size of the lattice $a(t)L$. For the sake of completeness we fit a broken linear function $0.5(1+(2\alpha)^n)^{1/n}$ to $\kappa(\alpha)$ and find a best fit value of the parameter $n=5.55$.

\begin{figure}
\centering
\includegraphics[width=0.49\textwidth]{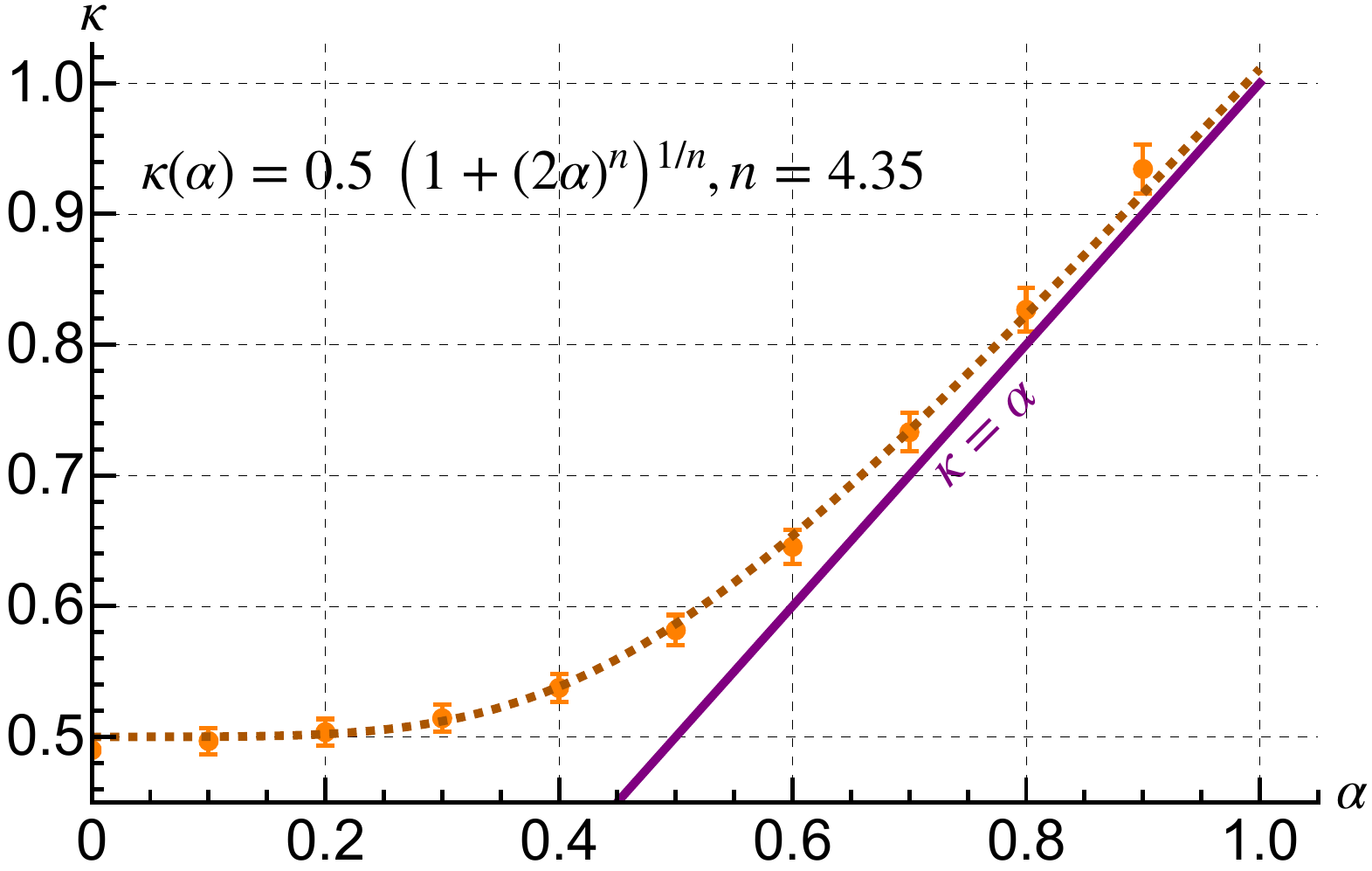}
\caption{\label{fig:dw_alpha_kappa} The late time power law index for $\mu_{\rm DW}(t)$, $\kappa$, given by $\mu_{\rm DW}(t)\sim t^{-\kappa}$,  as a function of the power law index, $\alpha$, associated with the growth of the scale factor (see Eq.~\eqref{eq:scfac}). The dashed orange line shows the broken linear function fit to the data points. For $\alpha \lesssim 1$ the dashed line has a slope of $\sim 1$, that is, $\kappa \sim \alpha$. The solid purple line shows $\kappa = \alpha$.  The error bars and the choice of parameters $L$, $N$, and $dt$ are discussed in Appendix~\ref{appsubsec:numerics_2}. 
}
\end{figure}

Similar results are obtained for the domain wall length density in two spatial dimensions. In particular at late times and for a given $\alpha$, the length density of domain walls, $\mu_{DW}$, also decays following a power law $t^{- \kappa}$. The associated annihilation process is harder to visualize since domain walls can have complex spatial structure in two dimensions and are no longer point-like. But it is not difficult to convince oneself that because they are created with random shapes, positions and velocities during the phase transition, and they percolate the lattice, at late times they are expected to disappear, and the whole lattice is expected to end up in one domain. We plot $\kappa$ with respect to $\alpha$ in Fig.~\ref{fig:dw_alpha_kappa}, and again notice a smooth interpolating behavior between $\kappa = 0.5$ at $\alpha=0$ (Minkowski spacetime) and $\kappa \sim \alpha$ for $\alpha\lesssim 1$. The latter case corresponds to a frozen network of domain walls blown away by the Hubble flow and whose length scales as $a(t)$. A broken linear function fit to $\kappa(\alpha)$ this time yields a best fit value of the parameter $n=4.35$. Because of our limited numerical precision, it is unclear whether the difference in parameters $n$ between the one- and two-dimensional cases is significant.

\section{VOS model}
\label{sec:VOS}

It is worth looking at our results from the perspective of the velocity-one-scale (VOS) model -- an effective description of the DW network in terms of two quantities: $L(t)$, the correlation length/average separation between  walls, and $v(t)$ the typical (root-mean-squared) wall velocity. 
For standard DW networks the VOS model was introduced in \cite{Avelino:2005kn} (see also \cite{Avelino:2022zem} and references therein).

As argued in  \cite{Pujolas:2022qvs},
VOS equations for a network of precursor DWs in an expanding universe should reduce to
\ba\label{VOS1}
\dot L &=&  H\, L + c \, v\,,   \\
\label{VOS2}
\dot v &=&  - k \frac{v^2}{L} - d \, H \, v   
\ea
where $d$ is the number of space dimensions, $H=\dot a/a = \alpha/t$ and $c,\, k$ the VOS parameters. The latter are expected to depend on $\alpha$ in general.  

Equations \eqref{VOS1}-\eqref{VOS2} have two special solutions. The trivial one,
\be\label{gas}
L(t) \propto a(t)\,, \qquad v=0\,,
\ee
corresponds to a `gas'  of noninteracting DWs. Wall-wall interactions are  frozen, and  the walls are carried and blown away by the expansion.

The other solution, for a power-law model $a(t)=\left(t/t_0\right)  ^\alpha$, is
\be
\begin{split}
L(t)&=  L_0 \left(t/t_0\right)^\kappa\,, \\
v(t)&= v_0 \, \left(t/t_0\right)^{\kappa-1} 
\end{split}
\label{curved}
\ee
with 
\ba\label{kappa_vv}
\kappa&=&\alpha + \vv \,,\\[2mm]
\vv &=&\frac{1-(d+1)\,\alpha}{1+k/c}~, \label{vv}
\ea
where we have introduced the `velocity parameter' as the combination 
\begin{equation}\label{vvDef}
    \vv \equiv \frac{v(t)}{L(t)/c\,t} =\frac{v_0}{L_0/c\,t_0} ~,
\end{equation}
which measures the typical velocity of the DWs in units of the characteristic velocity $L/c\,t$. (We introduce a conventional factor $c$ in \eqref{vvDef} for convenience). In contrast with the standard cosmological attractor of nonlinear DWs, here $L_0$ is a free parameter. The scaling \eqref{curved} is not characterized by a specific number of Hubble-sized DWs per Hubble patch, because this number increases steadily.

The parameter $\vv$ is more relevant. Physically it plays the role of a `temperature', the amount of DW motion. For any given $\alpha$ and $d$, it is in one-to one correspondence with (the ratio of) the VOS parameters $c,\,k$, see \eqref{kappa_vv}, so in a way it is a more physical parameterization. 
Importantly, it is clear from \eqref{kappa_vv} that $\vv$ also determines the scaling exponent $\kappa$, that is,  how $1/L$ and $v$ decay in time. This makes manifest that  the non-relativistic scaling of precursor DWs is tied to the  velocity distribution (in particular the velocity parameter $\vv$), as noted before in flat space~\cite{Pujolas:2022qvs}. From the VOS perspective, there is little more to be said without knowing $\vv$, that is, specifying the typical velocity $v$ in addition to the typical correlation length $L$.

\begin{figure}
\centering
\includegraphics[width=0.45\textwidth]{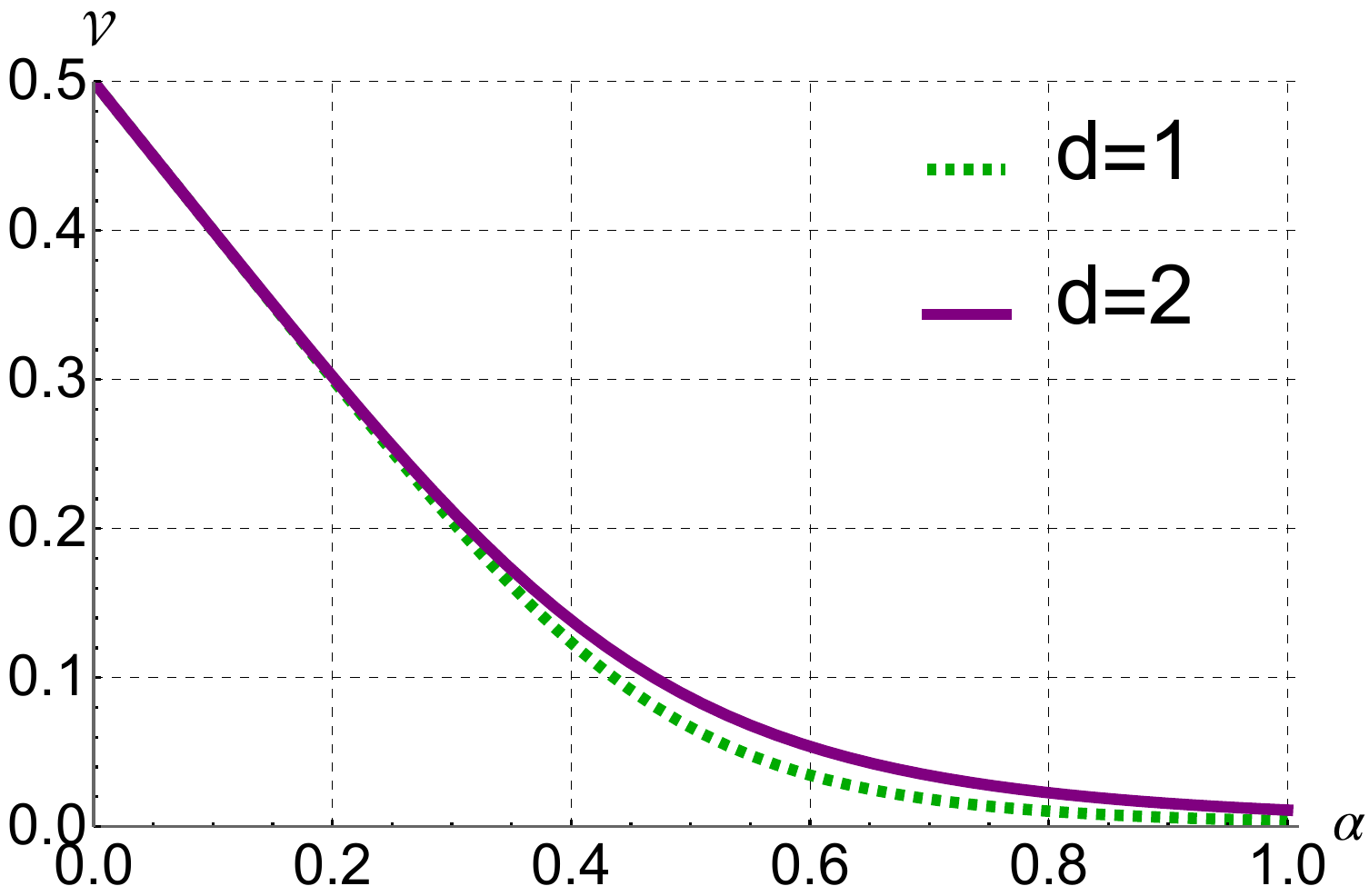}
\caption{\label{fig:v0alpha} 
Typical velocity in function of the expansion rate parameter $\alpha$ inferred from the VOS model. The vertical axis shows $v$ in units of $L/(c\, t)$. See Eq.~\eqref{vvDef}. 
}
\end{figure}

It would require considerable additional work to compute directly the distribution of velocities of kinks/DWs, or even their characteristic velocity. However, assuming the VOS equations \eqref{VOS1} and \eqref{VOS2}, we can already extract the velocity parameter $\vv$ from \eqref{kappa_vv}  together with the results shown in Figs.~\ref{fig:lambda_alpha} and \ref{fig:dw_alpha_kappa}\footnote{Similarly, using \eqref{kappa_vv} and \eqref{vv} we could read off how the ratio of VOS parameters $k/c$ depends on $\alpha$.}. The result is shown in Fig.~\ref{fig:v0alpha}. The 
nonrelativistic scaling ($\alpha\lesssim1/2$) - DW gas ($\alpha\gtrsim1/2$)
transition is manifest: 
$v(t)$ goes from an $O(1)$ fraction of the characteristic velocity for $\alpha\to0$ and drops almost linearly in $\alpha$ until $\alpha\sim1/2$ and then remains suppressed.

The interpretation of these results is clear.
In flat space, $\alpha=0$, $\vv$ coincides with the scaling exponent $\kappa$ and it takes the `universal' value\footnote{Equivalently, $k/c\to1$ for $\alpha\to0$} $\vv=1/2$, forced upon by non-the relativistic scaling that emerges at late times \cite{Pujolas:2022qvs}. 
The calculations in Secs. \ref{sec:kinks} and \ref{sec:dws} extend this to $\alpha\neq0$. The expansion forces the velocity parameter to decrease. However, for $\alpha \lesssim 1/2$, the correlation length exponent must remain close to $1/2$ because the correlation length is sub-Hubble -- the physical cosmic horizon grows faster than $L$ and so the effects of expansion must be small. In order for $\kappa(\alpha)$ to be approximately constant (as seen in Figs.~\ref{fig:lambda_alpha} and \ref{fig:dw_alpha_kappa}), the velocity parameter $\vv$ must be linear in $\alpha$. This explains the dependence  of $\kappa$ and $\vv$ on $\alpha$. At the core, for slow enough expansion ($\alpha\lesssim1/2$) the scaling is very close to the standard nonrelativistic one $L\sim t^{1/2}$. For faster expansion ($\alpha \sim 1/2$), this scaling is completely lost and the DWs are basically blown away with the Hubble flow.

\section{Discussion}
\label{sec:disc}
We studied the formation of domain walls (DWs) as a result of a cosmological $\mathbb{Z}_2$-symmetry breaking phase transition in an expanding background. Our description includes full quantum field theory effects but is limited to the regime of small fluctuations, that is to the precursor DWs regime. 
We find that the domain wall area density scales universally as $t^{-\kappa}$ at late times, independently of the details of the quench, with $\kappa$ depending only on the growth rate of the scale factor $\alpha=t\,H$ and the number of dimensions. Just like in the flat spacetime case, the quench time scale only impacts the domain wall area density at early times after the phase transition. 
The value of $\kappa$ in that case, $\alpha=0$, is known to be $1/2$ \cite{Mukhopadhyay:2020gwc,Mukhopadhyay:2020xmy}, which is interpreted as the wall network entering a nonrelativistic scaling regime \cite{Pujolas:2022qvs}. 
In presence of expansion, $\kappa$ depends on $\alpha$ in a distinctive way, interpolating smoothly between $\kappa\simeq 0.5$ for small growth rates $\alpha \lesssim 0.5$, and $\kappa\sim \alpha$ for $\alpha\gtrsim 0.5$. This behaviour is understood as follows. For $\alpha \lesssim 0.5$, the nonrelativistic scaling prevails (with small corrections) because the correlation length is always sub-Hubble and so one is `close' to the flat space case. However, for $\alpha > 1/2$ the scaling $L\sim t^{1/2}$ cannot be sustained, because the network is dynamical so $L$ is bound to grow at least as fast as the scale factor, $t^\alpha$.
Thus,  the expectation from Ref.~\cite{Pujolas:2022qvs} that there is a transition from nonrelativistic scaling to DW gas behaviour around  $\alpha\simeq0.5$ is confirmed. Moreover, we find the that the transition (the shape of $\kappa(\alpha)$) is smooth. 

Even though, for numerical reasons, our results are limited to low spatial dimensions, we expect qualitatively similar results to hold in the physically relevant three-dimensional case (and even in higher $d$). Indeed while the $1+1$ dimensional case is special because domain walls there have no internal degrees of freedom (they are point-like kinks and antikinks), already the two-dimensional case captures the richness of the continuously infinite number of degrees of freedom expected of domain walls in higher dimensions. 

In fact, a somewhat unexpected outcome of our work is that the details of the  (nonrelativistic scaling) - (DW gas) transition depend quite weakly on dimensionality, as is evident from the similitude of Figs. \ref{fig:lambda_alpha} (valid for $d=1$) and \ref{fig:dw_alpha_kappa} (valid for $d=2$). At first glance this seems very reasonable because DWs are codimension-1 objects, so a DW network in any number of dimensions should retain features of a bath of particles in $1+1$. However, DWs in $d+1$ interact/interconnect in much richer ways, so $O(1)$ discrepancies could have been present. In this sense, the most dramatic comparison is with the $d=1$ case where DWs are point-like. In other words, the closeness between the $d=1$ and $d=2$ cases suggests that $d=3$ should not differ greatly, because the major qualitative difference -- the extended character of the DWs -- is already present in $d=2$.

This suggests the following extrapolation of our results to $d=3$ as a reasonable expectation. One would expect $\kappa\to1/2$ for $\alpha \to0$, $\kappa\to\alpha$ for $\alpha \gtrsim1/2$. Concerning the most  interesting case phenomenologically, $\alpha=1/2$ or radiation domination, the tendency seems to be that $\kappa$ should depart from $1/2$, slightly increasing its value between $d=1$ and $d=2$. Assuming that this tendency is preserved, this points to a ballpark value of $\kappa\sim 0.6$. It would be interesting to see if this is confirmed.

Another interesting extension of our work is to include temperature. Thermal phase transitions have  been considered in the instantaneous quench limit~\cite{CALZETTA198932,Boyanovsky:1993fy,Ibaceta:1998yy,Boyanovsky:1999wd}. A network of precursor DWs is expected to  display the same `universal' late-time scaling behaviour derived here. It would be interesting to extend our calculation and check that this is indeed the case.

An important question which we haven't addressed is of course how feasible it is to have a precursor DW network epoch in realistic models. Is it possible that a DW network would spend a noticeable amount of time in an early scaling regime such as the one described here? This  requires very weak coupling or that the initial field amplitude at the quench be very small, which do not seem unreasonable conditions. Another possibility is that the precursor limit captures properties of 
standard, nonlinear DWs in friction-dominated models, as these fall in a non-relativistic regime too. We leave this for future research.

For the moment, let us point out that even if it turns out that a long period with scaling precursor DW networks cannot be realized in phenomenologically interesting models, precursor DWs still provide a nice and simple theoretical laboratory to understand the behaviour of extended objects in cosmology. Indeed, it is quite intriguing that interacting DW-like objects with rich `dynamics'
emerge from a free theory.
\acknowledgments
We thank Tanmay Vachaspati for useful discussions and comments. 
Computations for this research were performed on the Pennsylvania State University’s Institute for Computational and Data Sciences’ Roar supercomputer. We also acknowledge the use of the Deep Learning for Statistics, Astrophysics, Geoscience, Engineering, Meteorology and Atmospheric Science, Physical Sciences and Psychology (DL-SAGEMAPP) computing cluster at the Institute for Computational and Data Sciences (ICDS) at the Pennsylvania State University.
M.\,M. and G.\,Z. wish to thank the Yukawa Institute for Theoretical Physics, Kyoto University for hospitality during the period when this project was initiated.
M.\,M. is supported by National Science Foundation Grant No. AST-2108466. M.\,M. also acknowledges support from the 
Institute for Gravitation and the Cosmos Postdoctoral Fellowship at Pennsylvania State University. 
OP acknowledges the support from the Departament de Recerca i Universitats from Generalitat de Catalunya to the Grup de Recerca ‘Grup de Física Teòrica UAB/IFAE’ (2021 SGR 00649) and the Spanish Ministry of Science and Innovation (PID2020-115845GB-I00/AEI/10.13039/501100011033). IFAE is partially funded by the CERCA program of the Generalitat de Catalunya.
G.\,Z. is
supported by SONATA BIS grant 2023/50/E/ST2/00231 from the
Polish National Science Centre.
\appendix 
\section{Schr\"odinger picture quantization}
\label{appsec:derivation}
Below we outline the steps leading to Eq.~\eqref{eq:probden}. In other words we derive the Schr\"odinger picture quantum evolution of the wavefunctional describing the quantum state of the field $\phi$. In general this can only be done in a consistent way by regularizing the model on a discrete lattice.

Assuming periodic boundary conditions at spatial infinity, we start by doing an integration by parts on~\eqref{eq:action} thus rewriting the Lagrangian density as
\be
\label{eq:disclag1}
\mathcal{L} = \int dx \ a(t)\left[ \frac{1}{2} \dot{\phi}^2 + \frac{1}{2} a(t)^{-2} \phi \phi^{\prime \prime} 
- \frac{1}{2} m^2 (t)\phi^2 \right].
\ee
As described in the main text, we then discretize it on an $N$ point periodic lattice of size $L$ and lattice spacing $\ell = L/N$. For any given lattice point $i$ the value of the field $\phi$ at that point is denoted $\phi_i$. The periodicity of the lattice dictates, $\phi_{N+1} = \phi_1$ and $\phi_{0} = \phi_N$. We use the central finite difference approximation to lowest order for the spatial double derivative, that is
\be
\phi^{\prime \prime}_i \rightarrow \frac{\phi_{i-1} - 2 \phi_i + \phi_{i+1} }{\ell^2}.
\ee
With these prescriptions, the discretized Lagrangian density reads
\ba
\mathcal{L}_{\rm disc.} &=& a(t) \ell  \sum_{i=1}^N \left[ \frac{1}{2} \dot{\phi}_i^2 - \frac{1}{2 (a(t) \ell)^2} \phi_i  \right. \nn \\
 &&\times\big(\phi_{i-1}-   \left. 2 \phi_i + \phi_{i+1}  \big) - \frac{1}{2} m^2(t) \phi_i^2 \right].
\ea
In matrix notation the above equation can be rewritten more compactly as
\be
\label{eq:discL}
\mathcal{L}_{\rm disc.} = \frac{a(t)\ell}{2} \dot{\bm \phi}^T \dot{\bm \phi} - \frac{a(t)\ell}{2} \bm \phi^T \bm \Omega^2(t)  \bm \phi,
\ee
where, $\bm \phi \equiv \big( \phi_1, \phi_2,\dots, \phi_N \big)^T$, and $\bm \Omega^2(t)$ is given by
\be
\label{eq:omegasq}
[\bm{\Omega}^2(t)]_{ij} = 
\begin{cases}
+{2}/(a(t) \ell)^2+m^2(t),& i=j\\
-{1}/(a(t) \ell)^2,& i=j\pm1\ (\text{mod}\ N)\\
0,&\text{otherwise}.
\end{cases}
\ee
The full field theory has thus effectively been reduced to an $N$ particle quantum mechanics model whose quantum state (in the Schr\"odinger picture) is fully determined by a wavefunctional $\Psi(t;\bm{\phi})$ obeying the Schr\"odinger equation
\be
\label{eq:schro}
i \frac{\partial \Psi}{\partial t} = -\frac{1}{2 a(t)\ell} \sum_{i = 1}^N \frac{\partial^2 \Psi}{\partial \phi_i^2} + \frac{a(t)\ell}{2} \bm \phi^T \bm \Omega^2(t) \bm \phi\ \Psi\,.
\ee
The last remaining ingredient is specifying a reasonable initial state. We choose the field to be in the instantaneous ground state of the $N$ particle quantum mechanics model at time $t_0$: this choice corresponds to the zeroth order adiabatic vacuum in a field theoretic language.

Now, since our model is quadratic, the state will remain Gaussian as time evolves. Following~\cite{Mukhopadhyay:2020gwc,Mukhopadhyay:2020xmy,Pujolas:2022qvs}, we then write
\be
\label{eq:wavefunc}
\Psi (t; \bm \phi) = \mathcal{N}(t)\ {\rm exp} \left[ \frac{i a(t) \ell}{2} \bm \phi^T \bm M(t) \bm \phi \right],
\ee
where $\bm{M}(t)$ is a $N \times N$ complex symmetric matrix verifying $M(t_0)=i\bm{\Omega}^2(t_0)^{1/2}$ and $\mathcal{N}$ is a normalization factor. Plugging this {\it ansatz} in the Schr\"odinger equation~\eqref{eq:schro} we obtain the matrix differential equation
\be
\label{eq:meq}
\dot{\bm M}(t) + \bm M^2(t) + \bm \Omega^2(t) + H(t) \bm M(t) = 0.
\ee
With the help of the mode functions $c_n(t)$ introduced in Eqs.~\eqref{eq:cn},~\eqref{eq:cn0}, and~\eqref{eq:cndot0} the matrix elements of $\bm M(t)$ can then be written as,
\be
[\bm M(t)]_{jk} = \frac{1}{N} \sum_{n = 0}^{N-1} c_n(t)^{-1}\dot{c}_n(t) e^{2 i \pi n (j-k)/N}.
\ee
Here the fact that $c_n = c_{N-n}$ for $1 \leq n \leq N-1$ can be used to convince oneself that $\bm M(t)$ is indeed symmetric. Using the Wronskian constraint
\be
\label{eq:constraint}
c_n(t)^* \dot{c}_n(t) - c_n(t)\dot{c}_n(t)^* = \frac{i}{a(t) \ell},
\ee
which can be inferred from~\eqref{eq:cn},~\eqref{eq:cn0} and~\eqref{eq:cndot0}, the probability density functional $\mathcal{P} (t;\bm \phi) \equiv |\Psi(t;\bm \phi)|^2$ can finally be expressed in terms of the covariance matrix~\eqref{eq:green} as in Eq.~\eqref{eq:probden}. 
\section{Numerics}
\label{appsec:numerics}
In this appendix we give some details about the numerical equations and methods used. We also discuss the choice of parameters for our numerical integration method and its sensitivity thereon.

\subsection{Equivalent evolution equations}
\label{appsubsec:numerics_equations}
This section focuses on the derivation of the numerical equations we effectively use to compute the number density of kinks. Analogous equations can be obtained for domain walls.

The idea is to recast the differential equations~\eqref{eq:cn} in a form that is more amenable to numerical treatment. Indeed, the exponential growth of the $c_n$s corresponding to unstable modes quickly leads to very large numbers that exceed machine precision thus limiting the duration for which the numerical simulations are reasonably accurate. This problem can be circumvented by factoring out the leading exponential behavior (see~\cite{Mukhopadhyay:2020gwc}). More precisely, by writing $c_n(t) \equiv \rho_n(t) e^{i \theta_n(t)}$ and using~\eqref{eq:constraint} we find
\be
\rho_n^2 \dot{\theta}_n = \frac{1}{2 a(t)\ell},
\ee
which allows us to rewrite~\eqref{eq:cn} as
\ba
\label{eq:rhon}
&&\ddot{\rho}_n+H(t)\dot{\rho}_n+\left[ \frac{4}{(a(t)\ell)^2} {\rm sin}^2 \left( \frac{\pi n}{N} \right) + m^2(t) \right]\rho_n\nn\\
&&\hspace{3cm}=\frac{1}{4 (a(t)\ell)^2\rho_n^3}.
\ea
The corresponding initial conditions are
\ba
\rho_n (t_0) &=& 
\frac{1}{\sqrt{2a_{0}\ell}} 
\left [  \frac{4}{(a_0\ell)^2}\sin^2 \left ( \frac{\pi n}{N}\right ) + m^2(t_0) \right ]^{-1/4}\label{eq:rhon_init}\\
\dot{\rho}_n (t_0) &=& 0.
\label{eq:rhondot_init}
\ea
Noticing that, for $t>t_{\rm pt}$, the fastest growing unstable mode is the zero mode and that $\rho_0(t)\sim e^{mt}$ for large $t$, we are led to define $q(t)\equiv \ln(\rho_0(t))$ and $r_n(t)\equiv\rho_n(t)e^{-q_n(t)}$. From~\eqref{eq:rhon},~\eqref{eq:rhon_init} and~\eqref{eq:rhondot_init} we then obtain
\be
\label{eq:qeq}
\ddot{q}+\dot{q}^{2}+H(t) \dot{q}+m^2(t)=\frac{e^{-4q}}{4 (a(t)\ell)^{2}}\,,
\ee
with initial conditions
\ba
\label{eq:q_init}
q(t_{0})&=&\ln\left[ \frac{1}{\sqrt{2 a_0 \ell}}\Big( m^2 (t_{0}) \Big)^{-1/4}\right],\\
\label{eq:qdot_init}
\dot{q}(t_{0})&=&0,
\ea
and
\ba
\label{eq:req}
&&\ddot{r}_{n}+\left( 2\dot{q} + H(t) \right) \dot{r}_{n} \nonumber\\
&&\hspace{-1cm}+ \frac{4}{(a(t)\ell)^2}\left[\sin^2 \left ( \frac{\pi n}{N}\right )
+ \frac{e^{-4q}}{16}\left( 1-r_{n}^{-4} \right) \right] r_{n} = 0,
\ea
with initial conditions
\ba
\label{eq:r_init}
r_{n}(t_{0}) &=& \left[1+\frac{4}{(a_0\ell)^2m^2(t_0)}\sin^2 \left ( \frac{\pi n}{N}\right )\right]^{-1/4}\\
\label{eq:rdot_init}
\dot{r}_{n}(t_{0})&=&0.
\ea
By construction, Eqs.~\eqref{eq:qeq} and~\eqref{eq:req} do not suffer from the precision issues that plague the mode function evolution equations in their original form~\eqref{eq:cn} and allow for long integration times. 
\subsection{Miscellanea}
\label{appsubsec:numerics_2}
\begin{figure*}
\centering
\subfloat[(a)] {\includegraphics[width=0.49\textwidth]{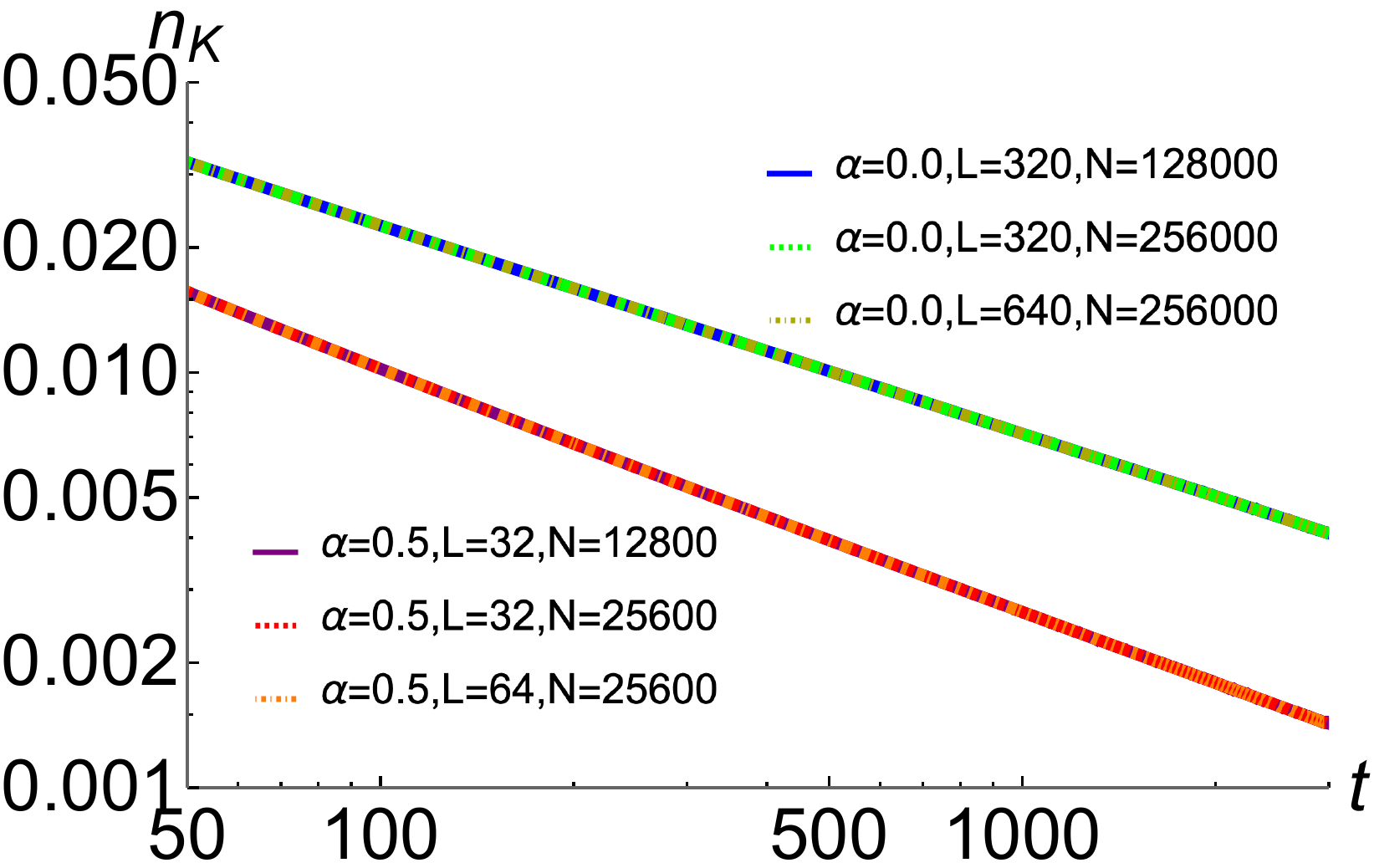}\label{fig:uv_ir_kinks}}\hfill
\subfloat[(c)] {\includegraphics[width=0.49\textwidth]{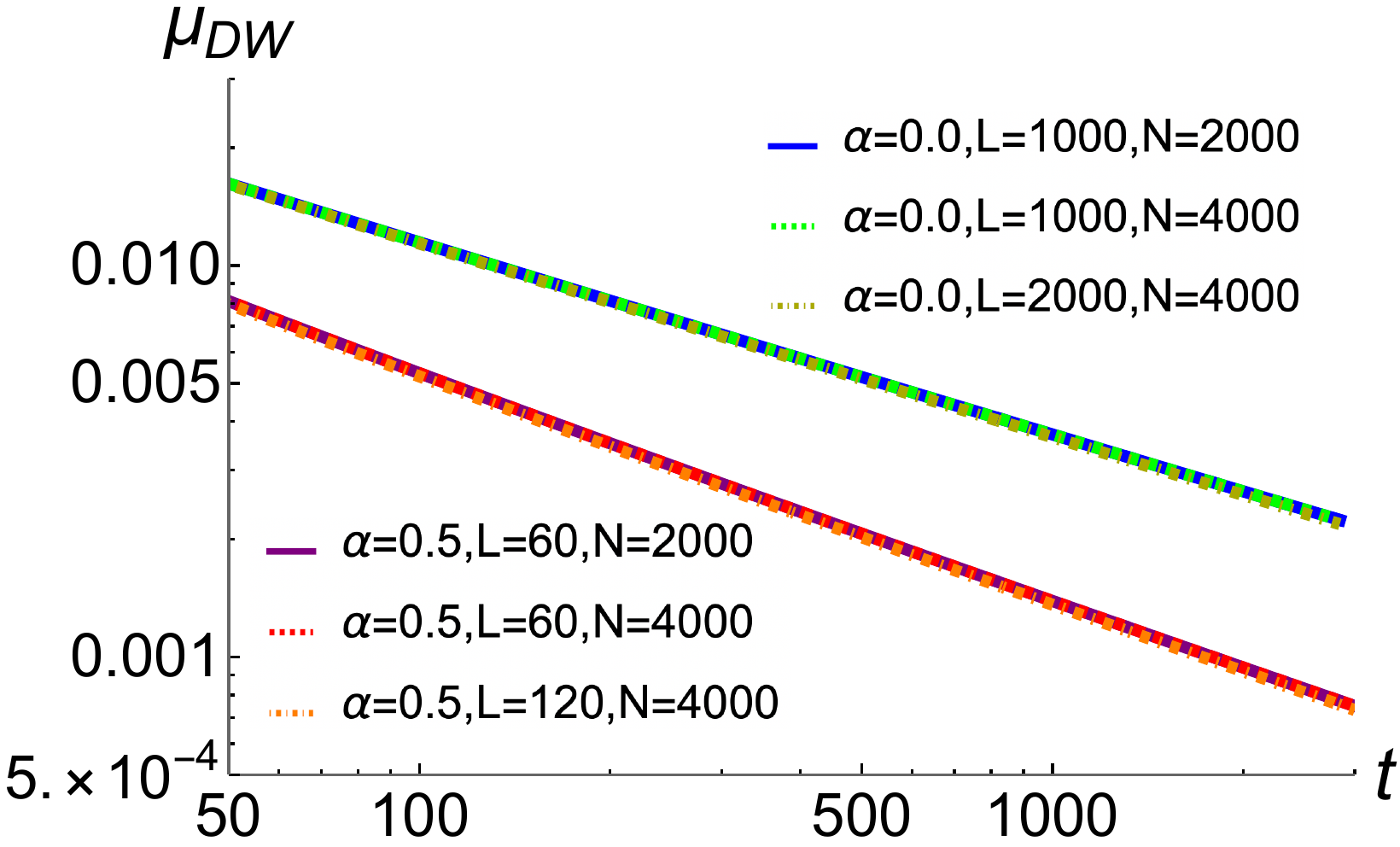}\label{fig:uv_ir_dws}}\hfill
\\
\subfloat[(b)] {\includegraphics[width=0.49\textwidth]{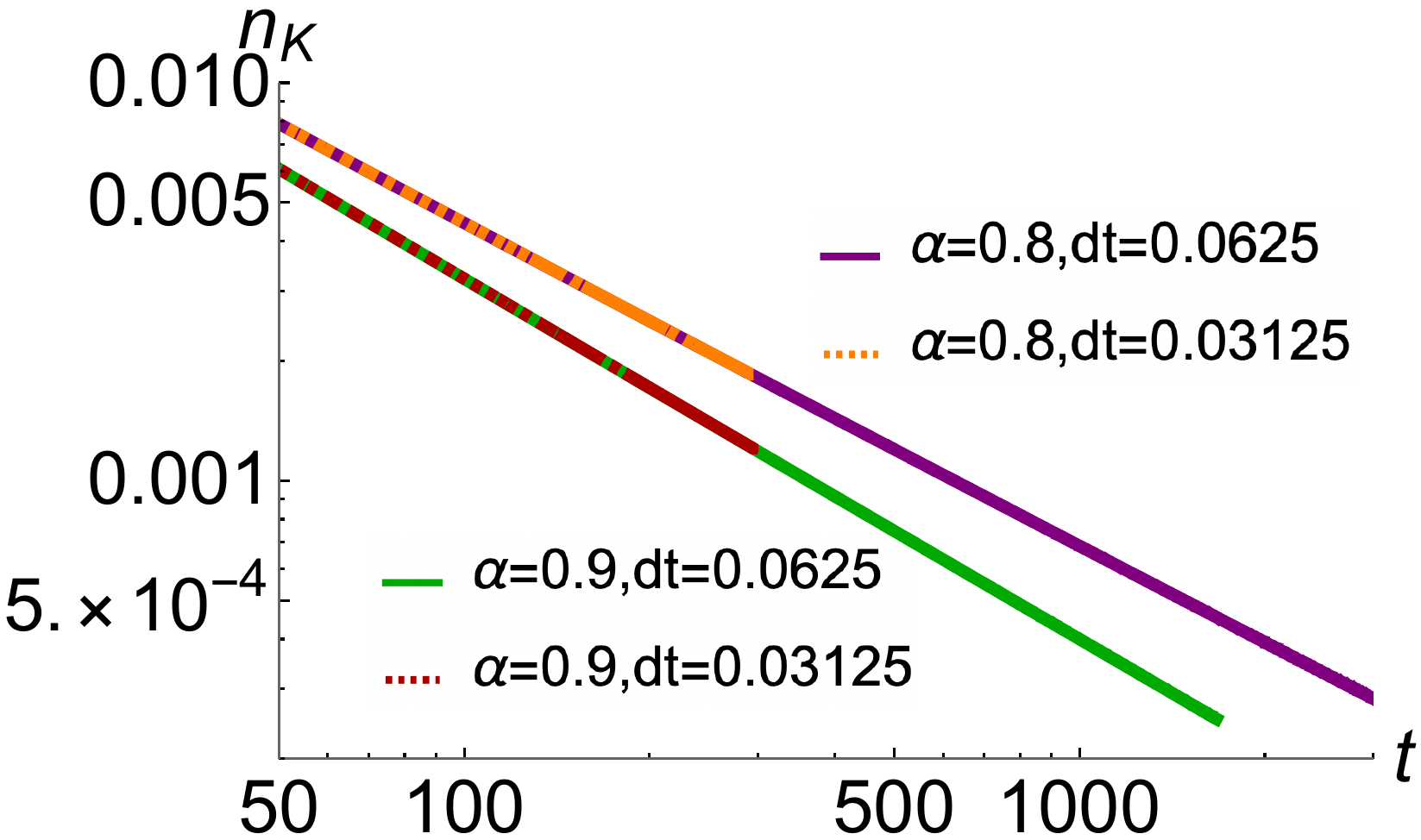}\label{fig:tcomp_kinks}} \hfill
\subfloat[(d)] {\includegraphics[width=0.49\textwidth]{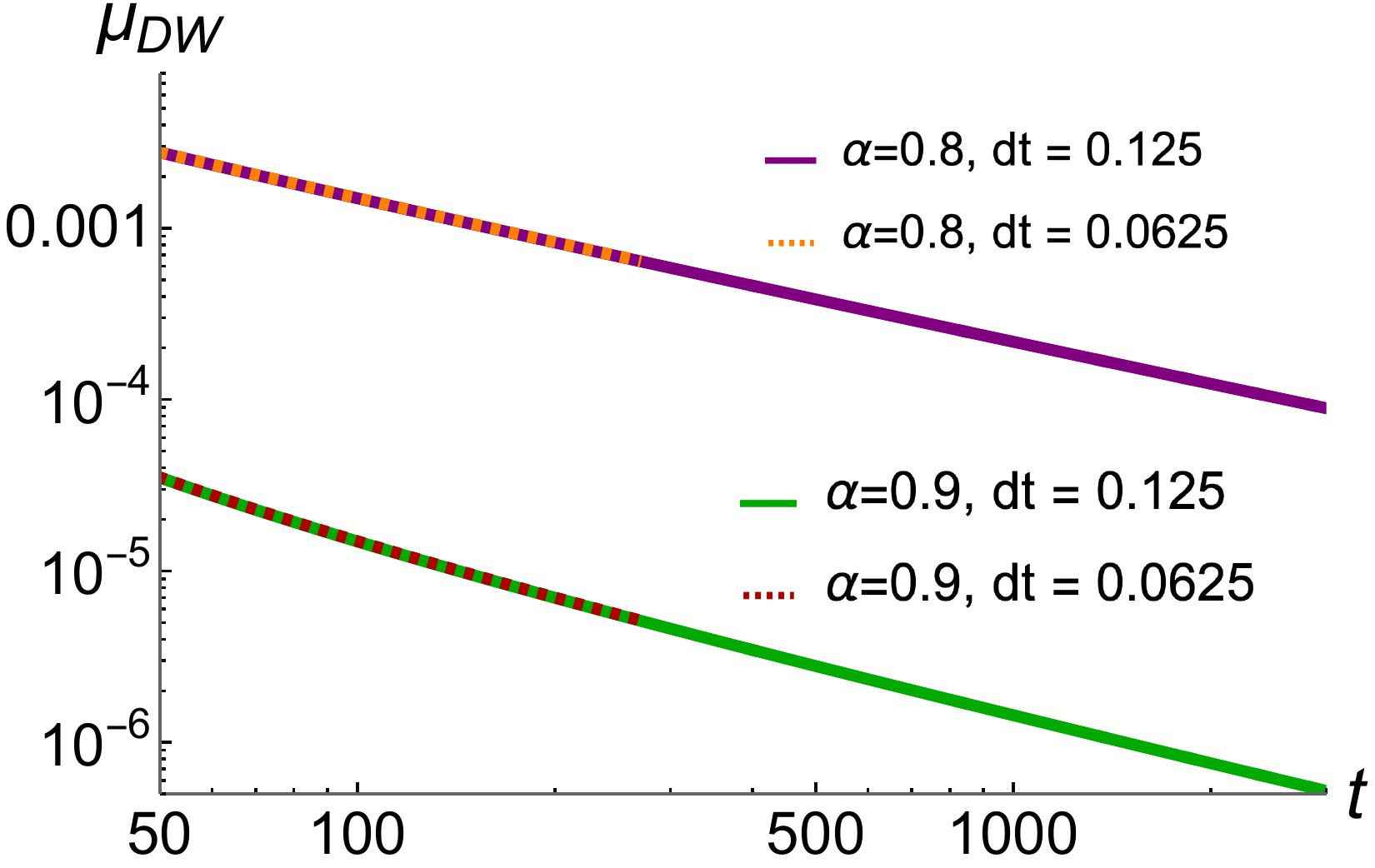}\label{fig:tcomp_dws}} \hfill
\caption{\label{fig:numerical_stuff} Number density of kinks ($n_K$, {\it left}) and length density of domain walls ($\mu_{DW}$, {\it right}) as a function of time, for different choices of the numerical evolution parameters $L$, $N$ (for $\alpha=0$ or $0.5$), and $dt$ (for $\alpha=0.8$ or $0.9$). The \emph{upper panels} illustrate the UV and IR cutoff insensitivity of the results, while the \emph{lower panels} illustrate their independence with respect to a finer choice of the temporal time step.
}
\end{figure*}
We now mention several subtle points that have an impact on how we carry out our numerical computations.

The fraction $u(t)$ of unstable modes in the system depends monotonically on $\ell t^\alpha$   as can be inferred from~\eqref{eq:ncrit}. To maintain accuracy, one needs to make sure that $u$ doesn't reach $\mathcal{O}(1)$ before the end of the numerical evolution. (In other words we need the unstable modes to not completely deplete the entire set of modes.) For a given total duration of the numerical evolution, this requires choosing a sufficiently small $\ell$. Moreover, the larger $\alpha$ is, the faster the expansion, and the smaller $\ell$ needs to be. This requirement is compatible with our focus on the continuum limit of the discretized system. Since we are ultimately interested in the infinite volume limit we also require $L$ (and consequently $N=L/\ell$) to be large.

An additional difficulty is related to the time increment $dt$ in our Crank-Nicholson numerical method. It turns out that the number of modes summed over in~\eqref{baralpha} and~\eqref{barbeta} becomes quite large for $\alpha>0.3$ and the numerical error in the computation of $c_n$ (depending on $dt$ and typically quite small) adds up to reach unacceptably large values if the time increment is not small enough. For faster expansion rates we need to substantially decrease $dt$ and therefore, in order to keep the runtime reasonably short, we are forced to decrease $N$ as well.\footnote{Since $u$ and $\ell$ are fixed by the desired evolution duration, $L$ needs to be smaller as well.}

For kinks in $1+1$ dimensions the computational complexity of the numerical method is $\mathcal{O}(N)$ and it is easy to satisfy the above conditions. We choose the lattice spacing to be $\ell = 0.0025$, which allows numerical evolution durations of order $10^3m^{-1}$ for even the largest $\alpha$s. This is achieved by choosing $L = 320, N = 128000$ for $\alpha \in \left[ 0.0,0.3 \right]$ and $L = 32, N = 12800$ for $\alpha \in \left( 0.3,0.9 \right]$. We choose $dt=0.5$ for $\alpha \in \left[ 0.0,0.3 \right]$, $dt=0.25$ for $\alpha \in \left( 0.3,0.7 \right]$, and $dt=0.0625$ for $\alpha \in \left( 0.7,0.9 \right]$. We see that as $\alpha$ increases we decrease $dt$, as well as $N$ and $L$, to keep numerical error under control. 

The above choices allow an accurate determination of the late time power law followed by the kink number density as a function of $\alpha$, as depicted in Fig.~\ref{fig:lambda_alpha}. However it turns out they are not sufficient for accurately capturing the immediate aftermath of the phase transition (where boundary effects are non-negligible for the previous choices of $L$). To achieve the desired precision (without having to increase $N$ as well) for very early times we perform shorter, lower-resolution runs (up to $t \sim 50m^{-1}$) with $L=3200, N=12800,$ and $dt=0.25$ (for all values of $\alpha$). We then combine the shorter runs with the main runs to obtain the smooth curves shown in Figs.~\ref{fig:numden_pow} and~\ref{fig:numden_tau_full}.

For domain walls in $2+1$ dimensions, numerical limitations place even greater constraints on our choices of parameters for the numerical evolution. Indeed the computational complexity is $\mathcal{O}(N^2)$ and we are effectively limited $N \lesssim \mathcal{O}(10^3)$. We choose $L = 1000$ and $N=2000$ for the static background ($\alpha = 0$). In this case we obtain $\kappa \approx 0.49$, a slightly different value from the exact one, $0.5$. This means our choice of $L$ and $N$ leads to higher uncertainty results for the values of $\kappa$, but it also provides a reasonable estimate for the expected error, $\sim 2\%$. This is encoded in the error bars around the points of Fig.~\ref{fig:dw_alpha_kappa}. We also need to contend with the same issues encountered in the case of kinks (and explained in the previous paragraphs). To ensure a numerical evolution duration of order $10^3m^{-1}$, we fix $N = 2000$ and decrease $L$ with increasing $\alpha$ as follows: $L = 1000$ for $\alpha = 0.1$, $L = 800$ for $\alpha = 0.2$, $L = 300$ for $\alpha = 0.3$, $L = 100$ for $\alpha = 0.4$, $L = 60$ for $\alpha = 0.5$, $L = 30$ for $\alpha = 0.6$, $L = 10$ for $\alpha = 0.7$, $L = 6$ for $\alpha = 0.8$, and $L = 2$  for $\alpha = 0.9$. The choice of such small $L$s is justified because the expansion increases the physical size of the box and we expect boundary effects to be negligible in the late time limit where the domain wall network is sufficiently sparse. To maintain nunmerical precision we also need to decrease the time increment with increasing $\alpha$ as follows: $dt = 0.5$ for $\alpha \in \left[0.0,0.7 \right)$, $dt = 0.25$ for $\alpha \in \left[0.7, 0.8 \right)$, and $dt = 0.125$ for $\alpha \in \left[0.8, 0.9 \right]$.

Finally, as a sanity check of the above parameter choices, in Fig.~\ref{fig:numerical_stuff}, we demonstrate the UV- and IR-independence of our results for two different values of $\alpha$ ($0$ and $0.5$). We also show that our choice of time evolution increment is sufficiently fine for the values of $\alpha$ most prone to numerical instabilities ($0.8$ and $0.9$). This proves the robustness of our results despite the computational limitations.
\bibliography{refs}
\bibstyle{aps}

\end{document}